\documentstyle[preprint,aps]{revtex}

\newcommand{\fseven}{1$f_{7/2}$}

\begin{document}
\draft

\title   {
            Shape Coexistence Around $^{44}_{16}$S$_{28}$:\\
             The Deformed  N=28 Region
         }
\author  {
           T.R. Werner,$^{a,b,}$\footnote{Institute of Theoretical
                            Physics, Warsaw University, Ho\.za 69,
                                            00-681 Warsaw, Poland},
           J.A. Sheikh,$^{c,}$\footnote{Tata
                            Institute of Fundamental Research, Colaba,
			    Bombay-400~005,  India},
           W. Nazarewicz,$^{a,b,c,*}$
           M.R.  Strayer,$^{a,b}$
           A.S. Umar,$^{d}$ \\
           and M. Misu$^b$
         }
\address {
           $^a$~Physics Division,  Oak Ridge National Laboratory, \\
                    P.O. Box 2008, Oak Ridge, Tennessee 37831, USA\\[1mm]
           $^b$~Department of Physics and Astronomy, University of Tennessee,\\
                                         Knoxville, Tennessee 37996, USA\\[1mm]
           $^c$~Joint Institute for Heavy-Ion Research, Oak Ridge,
	                                            Tennessee 37831, USA\\[1mm]
           $^d$~Department of Physics, Vanderbilt University,
                                               Nashville, Tennessee 37235,
USA\\
         }

\maketitle

\begin{abstract}
Masses, deformations, radii, and single-particle properties
of the very neutron-rich Sulfur isotopes are investigated
in the framework of the selfconsistent mean-field theory.
The stability of the
N=28 magic gap
around $^{44}$S is discussed.
\end{abstract}


\narrowtext
 In contrast to the nuclear structure along the beta-stability
line which has been well studied both experimentally and theoretically, the
yet unknown structure of drip-line nuclei
is currently of great interest \cite{Dourdan,[Roe92],[Mue93]}.
{}From the theoretical point of view, spectroscopy of exotic nuclei
offers a unique test of those components of effective interactions that
depend on isospin degrees of freedom.
 Since
the parameters of  interactions used in the usual mean-field
calculations
are determined so as to reproduce the
properties of beta-stable nuclei, these
 parameters may not always be optimal
around particle  drip lines due to (often
dramatic) extrapolations involved.

The nuclei discussed in this study are the Sulfur
isotopes, especially the neutron-rich ones  with N$\sim$28.
This particular choice was motivated
by recent experimental and astrophysical interest in this mass region
\cite{[Sor93],[Sch94],[Boe94]}.
{}From the perspective
of the spherical shell model, the underlying proton
wave functions  are described in terms of the $sd$
shell-model space,
while the main neutron components
originate from  the $fp$ shell-model space.
Such a schematic classification, however,  easily breaks down due
to the strong core polarization effect (i.e., the appearance
of static shape deformations
associated with the core-breaking excitations).
Experimentally,
deformed  states
in magic nuclei are known in many cases (see
the recent review \cite{[Woo92]}). Those {\em intruders}
sometimes appear very low in energy and, in a few cases,
they become ground states.

In the $sd$ region,
the neutron-rich nuclei with N$\sim$20 are spectacular examples of
coexistence between spherical and deformed configurations.
A classic example
of a magic nucleus with a deformed ground state
is $^{32}_{20}$Mg, which has a very low-lying
2$^+$ state at 886 keV \cite{[Det79]}
and an anomalously high value of $S_{2n}$.
Calculations based on the deformed mean-field theory predict deformed
ground states around $^{32}$Mg
and explain them in terms of neutron excitations to the
1$f_{7/2}$ shell \cite{[Cam75],[Ben84]}.
A similar conclusion has been drawn in the shell model calculations
in the ($sd$)($fp$) model space \cite{[Pov87],[War90a]} and in the
schematic analysis of ref. \cite{[Hey91a]}.

Another region of unexpected collectivity are the {\fseven} systems
around $^{48}_{24}$Cr$_{24}$. Naively, the main features of these
nuclei should be well reproduced in terms of a single-$j$ shell-model picture
governed by pairing interaction. However,
many properties of the observed states cannot be accounted for by
the results of the empirical
({\fseven})$^n$ shell model calculations \cite{[Kut78]} and the
extension to the full ($f,p$) configuration space is necessary
\cite{[Sek87],[Cau94]}.
Several nuclei around $^{48}$Cr exhibit rotation-like level spacings
\cite{[Cam90],[Cam93]} and the
self-consistent calculations \cite{[San75],[Jaq84],[Sai81]} yield
deformed ground states.
In particular, $^{48}$Cr is calculated
to be prolate-deformed with quadrupole deformation $\beta_2$=0.28
\cite{[Sai81]}.

Experimentally,
little is known about the neutron-rich nuclei around $^{44}$S.
Recently,
$\beta$-decay properties of $^{44}$S and $^{45-47}$Cl
have been studied in {\sc ganil} \cite{[Lew89],[Sor93]}.
The half-life of $^{44}$S was found to be $T_{1/2}$=123$\pm$10 ms.
(Observation of $^{42}$Si, $^{45,46}$P, $^{48}$S, and $^{51}$Cl
has been reported in ref. \cite{[Lew90]}.) The structure of exotic
neutron-rich nuclei with 10$<$Z$<$20 will soon be studied
at {\small GSI} using the fragmentation reaction
$^{48}$Ca on $^{9}$Be  at relativistic
energies ($v/c$$\approx$0.5) \cite{[Sch94]}.

Theoretical information on the light N$\approx$28 nuclei is also very
scarce.
The stability of highly neutron-rich Si isotopes was investigated
in ref. \cite{[Nay78]} in the spherical HF framework with various Skyrme
interactions.
The results of recent large-scale mass
calculations using the finite-range droplet model (FRDM) \cite{[Mol93]}
or the extended Thomas-Fermi with Strutinski-integral model (ETFSI)
\cite{[Abo92]} suggest the presence of deformation in some
N=28 isotones (see, e.g., Table IV of ref. \cite{[Sor93]}).
Based on these calculations and on the
theoretical analysis of $\beta$-decay properties of $^{44}$S,
it has been suggested in ref. \cite{[Sor93]} that
the influence of the spherical shell N=28  is weakened
below $^{48}$Ca.
The microscopic structure
of neutron-rich N$\sim$28 nuclei is important in
the astrophysical context.
Indeed, the neutron-rich N$\approx$28
nuclei are crucial for the nucleosynthesis of the
heavy Ca-Ti-Cr isotopes \cite{[Sor93]}.

To shed some light on the physics of exotic neutron-rich nuclei with
N$\approx$28, we performed calculations based on
the self-consistent mean-field theory,
namely the Skyrme-HF model
and
the relativistic mean-field (RMF) model.
In this Letter we report results for the Sulfur isotopes.
The results for the Silicon  and Argon isotopes will be published
in a forthcoming, more detailed paper.

We perform the Skyrme-HF calculations by discretizing
the energy functional on a three-dimensional Cartesian
spline collocation lattice, which provides a highly accurate alternative
to the finite-difference method \cite{[Uma91]}.
The structure of the resulting lattice representation is suited
for vector and parallel supercomputers, and
the method allows for highly modular programming where the
order of the splines can be defined as an input parameter.
Equations of motion are obtained
via the variation of the lattice representations of the constants of
motion, such as the total energy.
It is worth noting that no self-consistent symmetry has been
imposed in the calculations.
The details of our method have been published in Ref. \cite{[Uma91a]}.
In this work we use the Skyrme interaction
{\small SIII} \cite{[Bei75]}.
The exchange part of the Coulomb interaction
is taken to be
in the Slater form. In addition, we use the simple scaling
of the nuclear mass to approximately correct for the center-of-mass motion.
The calculations were performed in the cube of the size (20\ fm)$^3$.
In all the cases considered, the resulting HF minima turned out
to correspond to
reflection-symmetric shapes with three symmetry planes, i.e.,
$\langle x\rangle = \langle y\rangle = \langle z\rangle =0$ and
$\langle xy\rangle = \langle xz\rangle = \langle yz\rangle =0$.

The basic building blocks in the
relativistic mean-field
approach \cite{[Ser86]} are
the baryons (protons and neutrons) and the
 $\sigma -$, $\omega -$, and $\rho - $ mesons.
 The $\sigma$-meson is assumed to move in a non-linear
potential
\begin{equation}
U(\sigma) = {1\over 2} m_{\sigma} \sigma^2 + {1\over 3} g_2
\sigma^3
+{1\over 4} g_3 \sigma^4.
\end{equation}
In the present work we have employed the recent  set
 of Lagrangian density
parameters {\small NL-SH} ref.\cite{[Sha93]}.
This set has been claimed to be particularly
successful in describing properties of very neutron-rich
systems.
The relativistic
equations of motion are derived by means of the variational
principle. The resulting Dirac equation for the baryons
and  the Klein-Gordon equations for the mesons
are  solved
using the basis expansion
method.
In this method the small and the large components of the Dirac
spinor and the meson fields are expanded in terms of the
axially symmetric-stretched harmonic
oscillator basis with oscillator frequency $\hbar\omega$=41A$^{-1/3}$MeV.
(twelve deformed
oscillator shells for neutrons and protons were used). In the RMF
calculations, only reflection-symmetric axial shapes were considered.

It is known \cite{[Dob84]} that
in drip-line nuclei, pairing interaction leads to the
scattering of nucleonic pairs
from bound states to continuum.
In the mean-field+{BCS} model
this leads to the presence of unphysical
``particle gas" surrounding the nucleus.
The deficiencies of the standard treatment of pairing
around drip lines can be cured by means of the
Hartree-Fock-Bogolyubov (HFB)
method with a realistic
pairing interaction in which
the wave functions
of occupied quasiparticle states have correct asymptotic
behavior\,\cite{[Dob84]}. Unfortunately, the three-dimensional
{HFB} code is not available at present.
Consequently, in our study
we used the ``constant-gap"
approximation with a
strongly reduced pairing,
$\Delta_n$=$\Delta_p$=75 keV and 200 keV
in the {RMF} and {HF} models, respectively.
We also performed the RMF calculations with
$\Delta_n$=$\Delta_p$=500 keV. The results obtained were very similar to those
with $\Delta_n$=$\Delta_p$=75 keV.

The ground-state minima of the Sulfur  isotopes,
in most cases,
can be associated with
deformed intrinsic states. In order to compare
various variants of calculations and to relate them to previous
work, the standard quadrupole deformation parameters $\beta_2$
and $\gamma$  were extracted.
Firstly, the two quadrupole moments,
$q_{20}=\sqrt{16\pi/5}\langle r^2Y_{20}\rangle$
and $q_{22}=\sqrt{8\pi/15}\langle
r^2\left(Y_{22}+Y_{2-2}\right)\rangle$
were expressed in terms of the polar
coordinates $Q_\circ$ and $\gamma\ $\cite{[Hil53],[Boh52]}
\begin{equation}\label{quadr}
q_{20}=Q_\circ\mbox{cos}\gamma,
\hspace{1cm}
q_{22}=\frac{1}{\sqrt{3}}Q_\circ\mbox{sin}\gamma.
\end{equation}
Using Eq.~(\ref{quadr}), the $\gamma$--value  can be  determined.
The quadrupole moment of proton  distribution,
$Q_\circ^p$, can be written in terms of $\beta_2^p$ by means of
relation
\begin{equation}\label{quadr1}
Q_\circ^p=\sqrt{\frac{5}{\pi}}{\rm Z}\langle r^2_p\rangle \beta_2^p.
\end{equation}
Similarly, one can extract quadrupole deformations
of neutron ($\beta_2^n$) and
 mass ($\beta_2^{\rm A}$) distributions.
Since, in most cases, the equilibrium shapes predicted by
the {HF} model are axial, we adopted the standard convention (i.e.,
$\beta_2$$>$0 (prolate) for $\gamma$$\sim$0$^\circ$ and
$\beta_2$$<$0 (oblate) for $\gamma$$\sim$60$^\circ$).

The two-neutron separation energies
for the Sulfur isotopes, $S_{2n}$, are displayed
in Fig.~\ref{S2n} as a function of neutron number.
Our HF and RMF results are compared with
the predictions of the FRDM and ETFSI
models, and experimental data. In general, the agreement
between  the models themselves and between theory and experiment  is good.
In the HF   model the nucleus $^{52}$S
(N=36) is two-neutron-unstable. The RMF  calculations yield
more neutron binding;
the isotopes $^{52,54}$S are predicted to lie
inside the two-neutron drip-line.
The systematic model agreement for $S_{2n}$ should not imply that
the intrinsic structures of neutron-rich Sulfur
isotopes are also similar in all models presented in
Fig.~\ref{S2n}. As discussed below, this is not the case.

The calculated mass quadrupole deformations
of even-even Sulfur isotopes are shown
in Fig.~\ref{defs}  as a function of neutron number.
(Usually, calculations yield more than one energy minimum. In such
situations, deformations and excitation energies
 of excited states are also displayed.)
The deformation pattern for
even-even isotopes
$^{28-38}$S is fairly  similar in
the HF and RMF models: prolate ground states in  $^{28-32,38}$S,
oblate minimum in $^{34}$S, and spherical shape in the magic
N=20 nucleus
\renewcommand{\thefootnote}{\fnsymbol{footnote}}
\addtocounter{footnote}{2}
$^{36}$S\footnote{It is to be noted here
that RMF predicts a slightly deformed shape for
$^{36}$S. This can be attributed to very weak pairing in the
calculations.
 If a stronger proton pairing gap is assumed, $\Delta_p$=1 MeV,
the equilibrium shape becomes spherical.}.
However, the differences show up
for the heavier isotopes.

The nuclei $^{40,42}$S are predicted by the RMF model to have
prolate ground states with $\beta_2$$\sim$0.25;
the oblate minima
with $\beta_2$$\sim$-0.16
lie about 4 MeV higher in energy. In the HF
calculations,
prolate ($\beta_2$$\sim$0.25) and oblate
($\beta_2$$\sim$-0.24)
configurations
are practically degenerate.
The FRDM and ETFSI models give prolate-deformed ground states
with
($\beta_2$$\sim$0.25), in agreement with our findings.

There is no consensus  regarding  the equilibrium
shape of the N=28 nucleus  $^{44}$S.
According to  {RMF} calculations,  $^{44}$S
has  a  well-deformed prolate ground state with $\beta_2$=0.31. The
oblate minimum with $\beta_2$=0.16 appears to lie 0.8 MeV higher.
The spherical saddle point lies
even higher at $E^*$=2.8 MeV.
On the other hand, according to the {HF} model, $^{44}$S
is a gamma-soft system with  a small quadrupole deformation
$\beta_2$$\sim$0.13. The very shallow, well-deformed prolate minimum
($\beta_2$=0.28) analogous to the RMF ground state
lies at $E^*$$\sim$1.5\ MeV.
The {FRDM} and
{ETFSI} models give spherical and oblate
($\beta_2$=--0.26) ground states, respectively.

The heavier Sulfur isotopes
with N$>$28 are consistently calculated to be prolate-deformed in
the HF, RMF,
and FRDM
models.
The oblate minima are higher in energy, but
with increasing neutron number, the prolate-oblate
energy difference is reduced.
The RMF approach yields systematically larger quadrupole
deformations compared to HF. For instance, according to the RMF
predictions,  $^{48}$S is a well-deformed system with
$\beta_2$=0.22. According to the HF model, it is a transitional
nucleus with $\beta_2$=0.13, $\gamma$=10$^\circ$.
Interestingly, the {ETFSI} approach favors
a strongly deformed oblate configuration with $\beta_2$=--0.25.

For the N=28 isotone,
$^{42}$Si calculations yield well-deformed
oblate ground states with $\beta_2$ ranging
from --0.32 (FRDM) to --0.22 (HF). Also,  for the Silicon isotopes,
the RMF model gives slightly more neutron binding than
HF.
According to RMF, the nucleus $^{48}$Si is still inside the
two-neutron drip-line,
while it is unstable in the HF model.
Another
N=28 isotone,
$^{46}$Ar, is predicted to be spherical (RMF, FRDM) or oblate
(HF) with $\beta_2$=--0.14.
The RMF model gives an oblate minimum  with $\beta_2$=--0.10
at very low excitation energy. This indicates the transitional
character of $^{46}$Ar.

Figure\ \ref{sp1} displays the representative single-particle
neutron levels as functions of quadrupole deformation $\beta_2$. The
Nilsson diagram was obtained using a deformed Woods-Saxon model with
a Chepurnov set of parameters \cite{[Che67]}; the resulting
single-particle energies are close to those from the HF+{\small SIII}
model. (In the RMF--{\small NL-SH}  calculations
the order of $s_{1/2}$ and $d_{3/2}$ shells is reversed.)
 The neutron and proton single-particle levels in the ground-state of
$^{44}$S obtained in HF and RMF models are also displayed
in Fig.\ \ref{sp1}. The deformed shape in $^{44}$S results from
a subtle interplay between the deformed gaps at Z=16 and N=28,
and the spherical N=28 gap. In the RMF model, the
spherical N=28 gap is
completely broken;
the neutron intruder orbital [321]1/2,
originating from the $2p_{3/2,1/2}, 1f_{5/2}$ shells
above the N=28 gap  is occupied. In the HF calculations the
energy distance between the deformed
[321]1/2 and [303]7/2 levels is only 1.6 MeV.

The predicted root-mean-square (rms)  neutron and
charge radii of even-even  Sulfur
isotopes are illustrated in Fig.~\ref{radii}.
The results of HF and RMF models are fairly similar.
There appears  a systematic shift
of $\sim$0.08 fm between
the HF and RMF results  for the rms charge radii;
the experimental data
for $^{32,34,36}$S lie in between.
As expected, due to skin effect, rms radii increase
when approaching particle drip-lines.
(Since, in our study, the pairing correlations are practically neglected,
the undesired ``particle gas" effect mentioned above is not present.)

The difference between  neutron and proton deformations,
$\Delta\beta_2$$\equiv$$\beta_2^n-\beta_2^p$,
is illustrated in Fig.~\ref{defs1}. It is seen that when approaching
the neutron drip-line, the values of $\beta_2^n$ are systematically
smaller than those
 of the proton distribution. An opposite effect is
seen around the proton drip-line. The largest difference,
$|\Delta\beta_2|$$\sim$0.10, is obtained in the RMF model
for
$^{54}$S. (As discussed above, the deviation for $^{36}$S
disappears if the stronger proton pairing gap is assumed.)
The behavior of
$\Delta\beta_2$
can be partly attributed to the isotonic behavior of
rms radii in Fig.~\ref{radii} (see the definition
(\ref{quadr1}) of $\beta_2$).
 Indeed, the value of $Q_\circ$
depends both on the angular anisotropy
{\em and} the radial dependence of the nucleonic
density.
In the drip line nuclei, due to spatially extended wave functions,
the ``radial" contribution to $Q_\circ$ might be
as important as the ``angular" part. Moreover, it can strongly
depend on particle number.
The insert in Fig.~\ref{defs1}
illustrates the dependence of the $Q_\circ^n/Q_\circ^p$ ratio
on the N/Z ratio. Up to neutron number N=26 the HF and RMF results are very
similar. The dips in $Q_\circ^n/Q_\circ^p$ in the HF calculations
appear at shell and
subshell closures (i.e., at N=14, 20, 28, and 32). At these particle numbers
the neutron distribution has a tendency to be more spherical.
The lack of shell fluctuations in
$Q_\circ^n/Q_\circ^p$ in the RMF model at N$>$26
is consistent with the
calculated  large prolate deformations. Interestingly,
above N=28,  $\Delta\beta_2$(RMF) decreases steadily with
neutron number, while  the
$Q_\circ^n/Q_\circ^p$(RMF) ratio intersects the N/Z line
(rigid geometric
limit) only at N$>$34.

In summary, our calculations suggest strong deformation effects
in the region around $^{44}$S
due to the
{\fseven}$\rightarrow$$fp$
core breaking.
Large
differences between neutron and proton quadrupole
moments are obtained in the RMF approach
in deformed Sulfur nuclei far from stability. Such differences might have
interesting consequences for the quadrupole isovector
modes in drip line systems.
According to the RMF+{\small NL-SH} model, the N=28 nuclei $^{44}$S
and $^{42}$Si are
well deformed in their  ground states. On the other hand, the
HF+{\small SIII} model predicts smaller deformations, and
$^{44}$S is calculated to be a soft transitional system.
The new experimental data around $^{44}$S will certainly be very
helpful in pinning down the question of quadrupole collectivity
in this mass region.

Oak Ridge National
Laboratory is managed for the U.S. Department of Energy by Martin
 Marietta Energy Systems, Inc. under Contract No.
DE-AC05--84OR21400.
The Joint Institute for Heavy Ion
 Research has as member institutions the University of Tennessee,
Vanderbilt University, and the Oak Ridge National Laboratory; it
is supported by the members and by the Department of Energy
through Contract No. DE-FG05-87ER40361 with the University
of Tennessee.  Theoretical nuclear physics research
 is supported by the U.S. Department of
Energy through Contract No. DE-FG05-93ER40770 (University of Tennessee)
and through Contract No. DE-FG05-87ER40376 (Vanderbilt
University), and
by the Polish Committee for Scientific
Research through Contract No.{\ }20450~91~01.
In addition, this research project was
partially supported by the U.S. Department of Energy High Performance
Computing and Communications Program (HPCC) as the ``Quantum Structure of
Matter Grand Challenge'' project.
The numerical calculations were partially carried out on
the CRAY-2 supercomputer
at the National Energy Research Supercomputing Center,
Livermore.

\begin{figure}
\caption{Two-neutron separation energies
of the even-even Sulfur isotopes calculated with the
HF and RMF
models.
They are compared with the results of
the FRDM \protect\cite{[Mol93]}
and ETFSI
\protect\cite{[Abo92]} models, and experimental data.
 }
\label{S2n}
\end{figure}

\begin{figure}
\caption{
Quadrupole mass deformations $\beta_2^A$
of the even-even Sulfur  isotopes calculated with
the HF (top) and RMF (bottom) models.
The dots connected by a solid line correspond to  ground-state deformations.
The empty circles indicate excited configurations
(with excitation energies given in MeV).
}
\label{defs}
\end{figure}

\begin{figure}
\caption{Left: Woods-Saxon single-particle neutron levels
as functions of $\beta_2$.
The orbitals  are labeled
by means of $\Omega$ and $\pi$ ($\pi$=+, solid line; $\pi$=--, dashed line)
quantum numbers.
Right: The shell structure
of $^{44}$S as predicted by the HF and RMF models.
}
\label{sp1}
\end{figure}

\begin{figure}
\caption{Root-mean-square radii of the
neutron distribution (top) and charge distribution (bottom)
of the even-even Sulfur isotopes calculated with the
HF and RMF models.
The radii are shown
relative to the average (liquid-drop)
value of $\protect\sqrt{3/5}R_\circ$,
$R_\circ=1.2$A$^{1/3}$\ fm.
The experimental data for
$^{32,34,36}$S are taken from ref.\ \protect\cite{[Sch85]}
}
\label{radii}
\end{figure}

\begin{figure}
\caption{
Difference $\Delta\beta_2$$\equiv$$\beta_2^n-\beta_2^p$
for the even-even Sulfur
isotopes calculated with the
HF and RMF models as a function
of neutron number. The insert shows the ratio of neutron and proton
quadrupole moments versus N/Z. The large-deformation excited
state in $^{44}$S calculated in the HF model is indicated by means of
an ``($\bullet$)" symbol.
}
\label{defs1}
\end{figure}

\end{document}